\documentclass[twocolumn,showpacs,preprintnumbers,amsmath,amssymb,fleqn]{revtex4}
\usepackage{graphics}
\normalsize

\begin{document}

\title{Control of the entanglement of a two-level atom in a dissipative cavity via a classical field}

\author{Jian-Song Zhang} \author{Jing-Bo Xu}%
\email{ xujb@zju.edu.cn}

\address{Zhejiang Institute of Modern Physics and Physics Department,\\
Zhejiang University, Hangzhou 310027, People's Republic of China }

\begin{abstract}
{\bf Abstract}

 We investigate the entanglement dynamics and purity of a two-level
 atom, which is additionally driven by a classical field,
 interacting with a coherent field in a dissipative environment.
It is shown that the amount of entanglement and the purity of the
 system can be improved by controlling the classical
 field.

\pacs{03.67.Mn; 03.65.Ud}

Keywords: Cavity QED; Dissipation; Entanglement dynamics; Purity
\end{abstract}
\maketitle

\section{Introduction}
Quantum entanglement is at the heart of quantum information
processing and quantum computation \cite{Nielsen2000}. It can
exhibit a nonlocal correlation between quantum systems that can not
be accounted for classically. The cavity QED is a useful tool to
generate entangled states. It can be used to create entanglement
between atoms in cavities and establish quantum communications
between different optical cavities \cite{Raimond2001, Bose2001,
Kim2002, Li2005, Paternostro2007, Vitali2007, Rice2006}. Many
efforts have been devoted to the study of the manipulation of
quantum entanglement with atoms and photons in cavities. However, a
real quantum system is, in general, influenced by its surrounding
environment\cite{Paz2002}. The interaction between the quantum
system and its environment leads to the so-called
environment-induced decoherence. As a result, a pure quantum system
may then become mixed and the amount of its entanglement will
subsequently degrade.

In Ref.\cite{Solano2003}, Solano \emph{et al}.  have shown that
multipartite entanglement can be generated by putting several
two-level atoms in a cavity of high quality factor. In their paper,
the Schr\"{o}dinger cat state and other entangled states can be
produced with the help of a strong classical driving fields.
Generally, the entanglement of quantum states are fragile under the
influence of decoherence. This is the most serious problem for all
entanglement manipulation in quantum information processing. Up to
now, various methods have been proposed to suppress decoherence,
such as quantum error correction\cite{Shor1995}, decoherence-free
subspaces\cite{Zanardi1997}, quantum feedback
control\cite{Tombesi1995}, and dynamical decoupling\cite{Ban1998}.

In the present paper, we propose a scheme to enhance the amount of
entanglement and purity of a quantum system consisting of a
two-level atom interacting with coherent field in a dissipative
environment by applying and controlling a classical driving field.
We find an explicit expression of the density matrix of the system
by making use of the superoperator algebraic approach and study the
entanglement dynamics of the system by employing concurrence
\cite{Wootters1998}. Our calculation shows that the amount of
 entanglement and the purity of the system can be enhanced by controlling the
 classical driving
field.

\section{The model}

We consider a system consisting of a two-level atom interacting with
a coherent field. The atom is driven by a classical field
additionally. The Hamiltonian for the system can be described by
\cite{Solano2003}
\begin{eqnarray}
H&=&\omega a^{\dag}a+\frac{\omega_0}{2}\sigma_z+g(\sigma_+a
+\sigma_-a^{\dag})\nonumber\\
&&+\lambda(e^{-i\omega_ct}\sigma_++e^{i\omega_ct}\sigma_-),
\end{eqnarray}
where $\omega$, $\omega_0$ and $\omega_c$ are the frequency of the
cavity, atoms and classical field, respectively. The operators
$\sigma_z$ and
 $\sigma_{\pm}$ are defined by  $\sigma_{z}=|e\rangle\langle e|-|g\rangle\langle g|$ and
$\sigma_+=|e\rangle\langle g|$ where $|e\rangle$ and $|g\rangle$ are
the excited and ground states of the atom. Here, $a$ and $a^{\dag}$
are annihilation and creation operators of the cavity; g and
$\lambda$ are the coupling constants of the interactions of each
atom with the cavity and with the classical driving field,
respectively. Note that we have set $\hbar=1$ throughout this paper.

In the rotating reference frame the Hamiltonian of the system is
transformed to the Hamiltonian $H_1$ under a unitary transformation
$U_1= \exp{(-i\omega_c t\sigma_z/2)}$ \cite{Liu2006}
\begin{eqnarray}
H_1=U_1^{\dag}HU_1-iU_1^{\dag}\frac{\partial U_1}{\partial t}
=H_1^{(1)}+H_1^{(2)},
\end{eqnarray}
with
\begin{eqnarray}
H_1^{(1)}&=&\omega
a^{\dag}a+g(e^{i\omega_ct}\sigma_+a+e^{-i\omega_ct}\sigma_-a^{\dag}),\nonumber\\
H_1^{(2)}&=&\frac{\Delta_1}{2}\sigma_z+\lambda(\sigma_++\sigma_-),
\end{eqnarray}
and $\Delta_1=\omega_0-\omega_c$. A straightforward calculation
shows that the Hamiltonian $H_1^{(2)}$ can be diagonalized and
recast as
\begin{eqnarray}
H_1^{(2)}=\frac{\Omega_1}{2}\widetilde{\sigma}_z,
\end{eqnarray}
where $\Delta_1=\omega_0-\omega_c$,
$\Omega_1=\sqrt{\Delta_1^2+4\lambda^2}$ and $\widetilde{\sigma}_z$
is defined by $\widetilde{\sigma}_z=|0\rangle\langle
0|-|1\rangle\langle 1|$. Here, $|0\rangle$ and $|1\rangle$ are
dressed states
\begin{eqnarray}
|0\rangle=\cos{\frac{\theta}{2}}|e\rangle+\sin{\frac{\theta}{2}}|g\rangle,\quad
|1\rangle=-\sin{\frac{\theta}{2}}|e\rangle+\cos{\frac{\theta}{2}}|g\rangle,
\end{eqnarray}
with $\theta=\arctan{(\frac{2\lambda}{\Delta_1})}$. Neglecting the
terms which do not conserve energies (rotating wave approximation),
we obtain the effective Hamiltonian $H_1$ in the dressed states
\begin{eqnarray}
H_1&=&\omega a^{\dag}a+\frac{\Omega_1}{2}\widetilde{\sigma}_z
+g\cos^2{\frac{\theta}{2}}(e^{i\omega_ct}\widetilde{\sigma}_+a
+e^{-i\omega_ct}\widetilde{\sigma}_-a^{\dag}),\nonumber\\
\end{eqnarray}
with $\widetilde{\sigma}_+=|0\rangle\langle1|$ and
$\widetilde{\sigma}_-=|1\rangle\langle0|$. The Hamiltonian (6) can
be diagonalized by a final unitary transformation $U_2$ with
$U_2=\exp{(\frac{i\omega_ct}{2}\widetilde{\sigma}_z)}$. Using the
identity $e^{-i\omega_ct\widetilde{\sigma}_z/2}\widetilde{\sigma}_+
e^{i\omega_ct\widetilde{\sigma}_z/2}=e^{-i\omega_ct}\widetilde{\sigma}_+$,
we can rewrite the Hamiltonian of the system in the rotating
reference frame
\begin{eqnarray}
H_2&=&U_2^{\dag}HU_2-iU_2^{\dag}\frac{\partial U_2}{\partial t}\nonumber\\
&=&\omega a^{\dag}a+\frac{\omega'}{2}\widetilde{\sigma}_z+
g'(\widetilde{\sigma}_+a+\widetilde{\sigma}_- a^{\dag}),
\end{eqnarray}
where h.c stands for Hermitian conjugate,
$\omega'=\Omega_1+\omega_c=\sqrt{\Delta_1^2+4\lambda^2}+\omega_c$
and $g'=g \cos^2{\frac{\theta}{2}}$. It is worth noting that the
unitary transformations $U_1$ and $U_2$ are both local unitary
transformations. As we known the entanglement of a quantum system
does not change under local unitary transformations. Thus, the
entanglement of the system considered here will not change by
applying local unitary transformations $U_1$ and $U_2$. Hereafter,
unless specified otherwise we work in the rotating reference frame.

In the dispersive limit $|\Delta_2|=|\omega'-\omega|\gg\sqrt{n+1}g$,
the interaction Hamiltonian
$g'(a\widetilde{\sigma}_++a^{\dag}\widetilde{\sigma}_-)$ can be
regarded as a small perturbation. Using the method similar to that
used in Ref.\cite{Liu2006}, we can recast the effective Hamiltonian
(7) in the dispersive limit as
\begin{eqnarray}
H_e&=&\omega a^{\dag}a+\frac{\omega'}{2}\widetilde{\sigma}_z+\Omega
[(a^{\dag}a+1)|0\rangle\langle0|-a^{\dag}a|1\rangle\langle1|],
\end{eqnarray}
with $\Delta_2=\omega'-\omega$ and
$\Omega=\frac{(g\cos^2{\frac{\theta}{2}})^2}{\Delta_2}$.

\section{Solution}
In this section, we investigate the entanglement dynamics of the
two-level atom interacting with a coherent field in a dissipative
environment by making use of the superoperator algebraic approach
\cite{Xu1999}. We assume that a classical driving field is applied
additionally and
 the electromagnetic field couples to a reservoir. This
interaction causes the losses in the cavity which is presented by
the superoperator $\mathcal{D}=k(2a\cdot
a^{\dag}-a^{\dag}a\cdot-\cdot a^{\dag}a)$, where k is the decay
constant.  For the sake of simplicity, we confine our consideration
in the case of zero temperature cavity. In the interaction picture,
the interaction Hamiltonian is
\begin{eqnarray}
V=\Omega[(a^{\dag}a+1)|1\rangle\langle 1| -a^{\dag}a|1\rangle\langle
1|].
\end{eqnarray}
Then, the master equation that governs the dynamics of the system
can be written as follows
\begin{eqnarray}
\frac{d\rho}{dt}
&=&-i[V,\rho]+\mathcal{D}\rho\nonumber\\
&=&-i[V,\rho]+k(2a\rho
a^{\dag}-a^{\dag}a\rho-\rho a^{\dag}a).
\end{eqnarray}
We can express the density matrix in the following form
\begin{eqnarray}
\rho(t)&=&\rho_{00}(t)\otimes|0\rangle\langle0|+\rho_{11}(t)\otimes|1\rangle\langle1|
+\rho_{01}(t)\otimes|0\rangle\langle1|\nonumber\\
&&+\rho_{10}(t)\otimes|1\rangle\langle0|,
\end{eqnarray}
where $\rho_{ij}$'s are defined as $\rho_{ij}=\langle
i|\rho|j\rangle$, $\rho_{ij}=\rho_{ji}^{\dag}$, $i,j=0,1$. A
straightforward calculation shows that
\begin{eqnarray}
\dot{\rho}_{00}&=&\{-i\Omega(\mathcal{R}-\mathcal{L})+k(2\mathcal{M}-\mathcal{R}-\mathcal{L})\}\rho_{00}
=\mathcal{L}_{00}\rho_{00}(t),\nonumber\\
\dot{\rho}_{11}&=&\{i\Omega(\mathcal{R}-\mathcal{L})+k(2\mathcal{M}-\mathcal{R}-\mathcal{L})\}\rho_{00}
=\mathcal{L}_{11}\rho_{11}(t),\nonumber\\
\dot{\rho}_{01}&=&\{-i\Omega(\mathcal{R}+\mathcal{L}+1)+k(2\mathcal{M}-\mathcal{R}-\mathcal{L})\}\rho_{00}\nonumber\\
&=&\mathcal{L}_{01}\rho_{00}(t),\nonumber\\
\rho_{10}&=&\rho_{01}^{\dag}
\end{eqnarray}
where $\mathcal{M}$, $\mathcal{R}$, and $\mathcal{L}$ are defined by
\begin{eqnarray}
\mathcal{M}=a\cdot a^{\dag}, \mathcal{R}=a^{\dag}a\cdot,
\mathcal{L}=\cdot a^{\dag}a.
\end{eqnarray}
Here the superoperators $a\cdot, \cdot a, a^{\dag}\cdot$ and $\cdot
a^{\dag}$ represent the action of creation and annihilation
operators on an operator
\begin{eqnarray}
(a\cdot)o=ao, (\cdot a)o=oa,\quad (a^{\dag}\cdot)o=a^{\dag}o, (\cdot
a^{\dag})o=oa^{\dag}.
\end{eqnarray}
It is easy to check that the superoperators $\mathcal{M}$,
$\mathcal{R}$ and $\mathcal{L}$ satisfy the relations
\begin{eqnarray}
[\mathcal{R},\mathcal{M}]=[\mathcal{L},\mathcal{M}]=-\mathcal{M},
[\mathcal{R},\mathcal{L}]=0.
\end{eqnarray}
It is worth noting that
$[\mathcal{R}+\mathcal{L},\mathcal{M}]=-2\mathcal{M}$, the
superoperators $\mathcal{R}+\mathcal{L}$ and $\mathcal{M}$ form a
shift operator algebra. Thus we have the expansion of the
exponential of a linear combination of $\mathcal{R}+\mathcal{L}$ and
$\mathcal{M}$
\begin{eqnarray}
e^{\mathcal{L}_{00}t}&=&e^{(e^{2kt}-1)\mathcal{M}}e^{-(i\Omega+k)t\mathcal{R}}e^{(i\Omega-k)t\mathcal{L}},\nonumber\\
e^{\mathcal{L}_{11}t}&=&e^{(e^{2kt}-1)\mathcal{M}}e^{(i\Omega-k)t\mathcal{R}}e^{(-i\Omega+k)t\mathcal{L}},\\
e^{\mathcal{L}_{01}t}&=&e^{-i\Omega
t}e^{(e^{2(i\Omega+k)t}-1)\mathcal{M}/(i\Omega+k)}e^{-(i\Omega+k)t\mathcal{R}}e^{-(i\Omega+k)t\mathcal{L}}.\nonumber
\end{eqnarray}

We assume the field is initially prepared in the coherent state
$|\alpha\rangle$ and the atom is initially in state
$\frac{1}{\sqrt{2}}(|0\rangle+|1\rangle)$. Therefor the initial
state of the atom-cavity system is
$\frac{1}{\sqrt{2}}(|0\rangle+|1\rangle)\otimes|\alpha\rangle, $
i.e.,
$\rho_{00}(0)=\rho_{11}(0)=\rho_{01}(0)=\rho_{10}(0)=\frac{1}{2}|\alpha\rangle\langle\alpha|$.
Combing Eq.(12) with Eq.(16), we find that the matrix elements
$\rho_{ij}(t)$ at time t is given by
\begin{eqnarray}
\rho_{00}(t)&=&\frac{1}{2}|\alpha_+(t)\rangle\langle\alpha_+(t)|,\quad
\rho_{11}(t)=\frac{1}{2}|\alpha_-(t)\rangle\langle\alpha_-(t)|,\nonumber\\
\rho_{01}(t)&=&\frac{1}{2}f(t)|\alpha_+(t)\rangle\langle\alpha_-(t)|,
|\alpha_\pm(t)\rangle=|\alpha e^{-(k\pm i\Omega)t}\rangle,\nonumber\\
f(t)&=&\exp{\{-i\Omega
t+|\alpha|^2(e^{-2kt}-1)\}}\nonumber\\
&&\times
\exp{\{\frac{|\alpha|^2k}{k+i\Omega}(1-e^{-2(k+i\Omega)t})\}}.
\end{eqnarray}
The density matrix of the atom-field system is then expressed as
\begin{eqnarray}
\rho(t)&=&\frac{1}{2}\{|\alpha_+(t)\rangle\langle\alpha_+(t)|\otimes|0\rangle\langle
0|+|\alpha_-(t)\rangle\langle\alpha_-(t)|\\
&&\otimes|1\rangle\langle
1|+[f(t)|\alpha_+(t)\rangle\langle\alpha_-(t)|\otimes|0\rangle\langle
1|+h.c]\},\nonumber
\end{eqnarray}
where h.c stands for Hermitian conjugate. In order to map the matrix
onto a $2\times 2$ system, we introduce two orthonormal vectors
$|\uparrow\rangle$ and $|\downarrow\rangle$ which are defined by
\begin{eqnarray}
|\uparrow\rangle=|\alpha_+(t)\rangle,
|\downarrow\rangle=\frac{1}{\sqrt{1-|\tau|^2}}(|\alpha_-(t)\rangle-\tau|\alpha_+(t)\rangle),
\end{eqnarray}
with $\tau=\langle\alpha_+(t)|\alpha_-(t)\rangle$. Finally, the
density matrix $\rho(t)$ now can be rewritten as
\begin{eqnarray}
\rho(t)&=&\frac{1}{2}\{|\uparrow\rangle\langle\uparrow|\otimes|0\rangle\langle
0|+(\tau|\uparrow\rangle+\sqrt{1-|\tau|^2}|\downarrow\rangle)\nonumber\\
&&(\tau^*\langle\uparrow|+\sqrt{1-|\tau|^2}\langle\downarrow|)\otimes|1\rangle\langle1|
\nonumber\\
&&+[f(t)|\uparrow\rangle(\tau^*\langle\uparrow|+\sqrt{1-|\tau|^2}\langle\downarrow|)
\otimes|0\rangle\langle1|+h.c]\},\nonumber\\
\end{eqnarray}
where $h.c$ denotes for Hermitian conjugate.

\section{Entanglement and purity}
In this section, we investigate the entanglement dynamics and purity
of a quantum system consisting of a two-level
 atom, which is additionally driven by a classical field,
 interacting with a coherent field in a dissipative environment.

In order to study the entanglement of above system described by
density matrix $\rho$, we adopt the measure concurrence which is
defined by \cite{Wootters1998}
\begin{equation}
C=\max{\{0, \lambda_1-\lambda_2-\lambda_3-\lambda_4\}},
\end{equation}
where the $\lambda_i$(i=1,2,3,4) are the square roots of the
eigenvalues in decreasing order of the magnitude of the
``spin-flipped" density matrix operator
$R=\rho(\sigma_y\otimes\sigma_y)\rho^*(\sigma_y\otimes\sigma_y)$ and
$\sigma_y$ is the Pauli Y matrix, i.e., $\sigma_y= \left(
\begin{array}{cc}
  0 & -i \\
  i & 0
  \end{array}
  \right)$.
Combing the definition of concurrence with the density matrix, we
find that the concurrence of the system is
\begin{eqnarray}
C(t)=|f(t)|\sqrt{1-|\tau|^2}.
\end{eqnarray}

\begin{figure}
\centering { \scalebox{1}[1]{\includegraphics{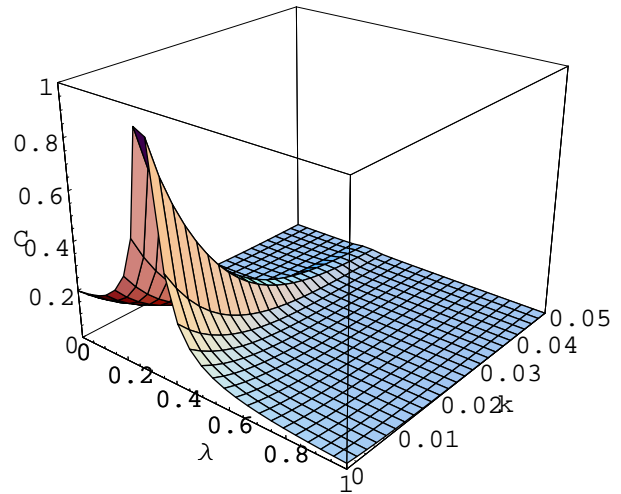}}} \caption{The
concurrence is plotted as a function of the coupling strength
$\lambda$ and the decay rate k with $\alpha=1$, $g=10^{-2}, t=1/g,
\omega=2, \omega_0=1.9, \omega_c=0$.}
\end{figure}

\begin{figure}
\centering { \scalebox{1}[1]{\includegraphics{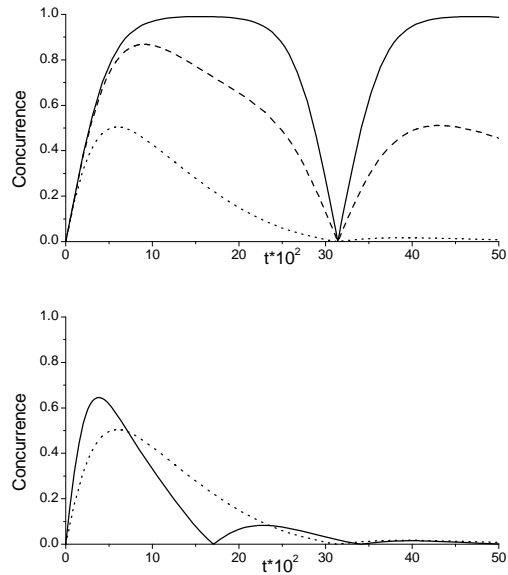}}}
\caption{Upper panel: The concurrence is plotted as a function of
time with $\alpha=1$, $g=10^{-2}, \omega=2, \omega_0=1.9,
\omega_c=\lambda=0$ for $k=0$ (solid line), $k=10^{-4}$ (dashed
line), and $k=10^{-3}$ (dotted line). Lower panel: The concurrence
is plotted as a function of time with $\alpha=1$,  $g=10^{-2},
\omega=2, \omega_0=1.9, k=10^{-3}$ for $\omega_c=\lambda=0$ (dotted
line) and $\omega_c=\lambda=0.2$ (solid line).}
\end{figure}

In order to show the the  effect of the classical field and the
decay of the cavity on the entanglement dynamics of the system, we
plot the the concurrence as a function of the coupling strength
$\lambda$ and the decay rate k in Fig.1. It is easy to see that the
entanglement of the two-level atom and the cavity decreases with the
increase of the decay rate k. However, the amount of entanglement
between the two-level atom and the cavity can be increased by
controlling the classical driving field as we can see from Fig.1.

In Fig.2, we plot the concurrence $C(t)$ as a function of time for
different values $k$, $\omega_0$, and $\lambda$. From the upper
panel of Fig.2, one can see that the entanglement of the system
decreases with the increase of the decay rate $k$. However, the
maximal value of entanglement for the system can be improved by
applying the classical driving field as one can easily find out in
the lower panel of Fig.2.

It is worth noting that in the case of $t\rightarrow \infty$ the
states $|\alpha_\pm(t)\rangle=|\alpha e^{-(k\pm i\Omega)t}\rangle$
eventually goes to vacuum state $|0\rangle$. The photon number of
the cavity field is
\begin{eqnarray}
\langle n\rangle=Tr\{\rho(t) a^{\dag}a\}=|\alpha|^2 e^{-2kt}.
\end{eqnarray}
It is easy to see that the photon number of the cavity field depends
only on the photon number of the initial state $|\alpha|^2$, the
decay rate k, and the time t. In Fig.3, we plot the photon number of
the cavity field as a function of time t with $\alpha=1$ for
$k=10^{-4}$ (dotted line) and $k=10^{-3}$ (solid line). Comparing
Fig.3 with Fig.4, we find that the photon number of the cavity field
decreases with the increase of the time t while the concurrence is
not a monotonic function of time t.

\begin{figure}
\centering { \scalebox{1}[1]{\includegraphics{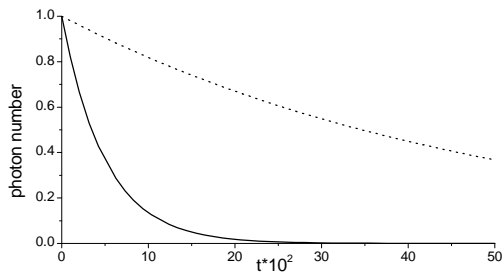}}} \caption{The
photon number of the cavity field is plotted as a function of time t
with $\alpha=1$ for $k=10^{-4}$ (dotted line) and $k=10^{-3}$ (solid
line). }
\end{figure}

Next, we investigate the purity of the system by employing linear
entropy. Many protocols in quantum information processing require
pure, maximally entangled quantum states. For example, quantum
teleportation often relies heavily on the purity and entanglement of
the initial state. However,  an pure and entangled quantum system
usually becomes mixed and/or less entangled under the influence of
decoherence. Here, we adopt the linear entropy to quantify the
mixedness of a state defined by $ S(\rho)=1-Tr(\rho^2).$ Generally,
if $\rho$ is the density matrix of a pure state, $S=0$, otherwise
$S>0$. It has also been proved that a bipartite mixed states is
useless for quantum teleportation if its linear entropy exceeds
$1/2$ for a two qubits system. The purity of the atom-field system
is
\begin{eqnarray}
S(\rho)=1-Tr(\rho^2)=\frac{1}{2}[1-|f(t)|^2].
\end{eqnarray}
In Fig.4, the linear entropy $S(\rho)$ is plotted as a function of
time with (solid line) or without (dotted line) the classical
driving field. As one can see clearly from Fig.4, the classical
driving field can decrease the linear entropy of the atom-field
system. In other words, the purity of the atom-field system can be
significantly increased by applying the classical driving field.

\begin{figure}
\centering { \scalebox{1.5}[1.5]{\includegraphics{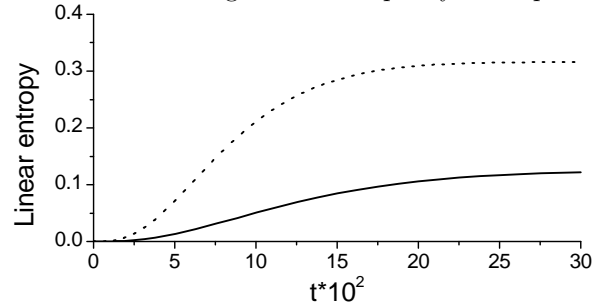}}} \caption{
The concurrence is plotted as a function of time with $\alpha=1$,
$g=10^{-2},\omega=2, \omega_0=1.9, k=10^{-3}$ for
$\omega_c=\lambda=0$ (dotted line) and $\omega_c=\lambda=0.5$ (solid
line).}
\end{figure}

\section{Conclusion}
In the present paper, we propose a scheme to improve the amount of
entanglement and purity of a quantum system consisting of a
two-level atom interacting with a coherent field in a dissipative
cavity by applying and controlling a classical driving field. We
find an explicit expression of the density matrix of the system and
study the entanglement dynamics of the system by employing
concurrence. Our calculation shows that the amount of entanglement
and the purity of the system can be enhanced by applying the
classical driving field. The approach adopted here can be extended
to the system formed by two or more two-level atoms in dissipative
cavities.

\section*{Acknowledgments}
 This project was supported by the National Natural
Science Foundation of China (Grant no.10774131) and the National Key
Project for Fundamental Research of China (Grant no. 2006CB921403).

\end{document}